\documentclass[showpacs,amsmath,amssymb,prb]{revtex4}
\usepackage{graphicx}
\usepackage{dcolumn}
\usepackage{bm}
\def\etal{\mbox{\it et al.\ }}

\begin{document}

\title{Compressibility effects on the Rayleigh-Taylor instability growth
between immiscible fluids}

\author{D. Livescu}
\affiliation{University of California,\\ 
Los Alamos National Laboratory,\\
Theoretical Division, T3 MS B216, Los Alamos NM 87545\\
livescu@lanl.gov}
\date{\today}

\begin{abstract}
The linearized Navier-Stokes equations for a system of superposed immiscible
compressible ideal fluids are analyzed. The results of the analysis reconcile the 
stabilizing and destabilizing effects of compressibility reported in the literature.
It is shown that the growth rate $n$ obtained for an inviscid, compressible flow in an 
infinite domain is bounded by the growth rates obtained for
the corresponding incompressible flows with uniform and exponentially varying 
density. As the equilibrium pressure at the interface $p_\infty$ increases (less 
compressible flow), $n$ increases towards the uniform density result, while as the 
ratio of specific heats $\gamma$ increases (less compressible fluid), $n$ decreases
towards the exponentially varying density incompressible flow result. This remains 
valid in the presence of surface tension or for viscous fluids and the validity of
the results is also discussed for finite size domains. The critical wavenumber 
imposed by the presence of surface tension is unaffected by compressibility. 
However, the results show that the surface tension modifies the sensitivity of the
 growth rate to a differential 
change in $\gamma$ for the lower and upper fluids. For the viscous case, the 
linearized equations are solved numerically for different values of $p_\infty$ and
$\gamma$. It is found that the largest differences compared with the incompressible
cases are obtained at small Atwood numbers. The most unstable mode for the 
compressible case is also bounded by the most unstable modes corresponding to the 
two limiting incompressible cases.

\end{abstract}


\maketitle

\section{Introduction
\label{intro}}

The Rayleigh-Taylor instability, which occurs due to the gravitational instability
of a heavy fluid overlying a lighter fluid, \cite{CH81,Taylor50} is of fundamental
importance in a multitude of applications ranging from the turbulent mixing in 
Inertial Confinement Fusion\cite{K94,Knauer00} to astrophysical phenomena. 
\cite{R94,FL00} Small perturbations of the interface between the two fluids grow 
to large amplitudes. At early times, for small enough initial perturbations, the 
flow can be described by the linearized equations and the amplitude grows 
exponentially. Later, the interface evolves into bubbles of lighter fluid and 
spikes of heavier fluid penetrating the opposed fluid. If the initial interface is
randomly perturbed then bubbles and spikes of different sizes are generated. 
Mathematically, the Rayleigh-Taylor instability is an ill-posed problem and the 
dependence on initial conditions is still an interesting question. \cite{Sharp84}

The theory for the linear stage for incompressible fluids agrees well with the
experiments. \cite{CH81,WNJ01} In the absence of surface tension and viscosity, the 
growth rate increases indefinitely with the wavenumber. This trend is changed 
by the presence of viscosity, in which case the growth rate has a peak value 
and decreases towards zero for large wavenumbers. On the other hand, the 
presence of surface tension stabilizes perturbations with the wavenumber larger
than a critical value. 

The role of compressibility on the development of the Rayleigh-Taylor instability 
between inviscid fluids has been studied by several authors, 
\cite{Sharp84,Vandervort61,PP82,Baker83,BB83,LP89,YZ93,Amala95,Turner02} however its 
effect compared with the incompressible case is still under debate. The earlier 
studies of the linear stage dealing with ideal fluids \cite{Vandervort61,PP82} 
introduced simplifying assumptions and are strictly valid only when $\gamma=1$. 
Bernstein and Book \cite{BB83} and Turner \cite{Turner02} removed these assumptions
and studied the effects of compressibility as a function of $\gamma$. They found 
that these effects are more important at small wavenumbers and the rate of growth 
increases as $\gamma$ decreases. They concluded that compressibility has a 
destabilizing effect. The same conclusion is obtained for a multilayer system by 
Yang and Zhang \cite{YZ93} by comparing the compressible growth rate with that 
corresponding to an incompressible system obtained as $\gamma \rightarrow
\infty$. The increase in the growth rate as $\gamma$ decreases can also be 
explained using the energy principle, as a special case of the comparison theorem 
in the calculus of variations. \cite{Newcomb83}

On the other hand, Sharp \cite{Sharp84} finds a stabilizing effect of 
compressibility. Moreover, numerical results for late time growth  seem to indicate
that $\alpha$ (the constant of proportionality in the quadratic law for the rate of
growth) increases with the speed of sound, \cite{Li93} so that compressibility 
would have a stabilizing effect. Baker \cite{Baker83} found both stabilizing and 
destabilizing effects of compressibility in the linear regime, however his results
were based on previously derived formulas using different assumptions than ideal 
gas. The role of compressibility on the instability growth is thus not yet settled.
Moreover, to the best of our knowledge no study of the effects of compressibility 
in the linear regime for viscous fluids has been performed. Additionally, there
is no systematic study of the effects of surface tension and finite size domain 
in the compressible case. 

In this paper we resolve the apparent contradiction between the stabilizing and
destabilizing character of compressibility for ideal fluids. We show that 
compressibility can be characterized by two parameters, $\gamma$ and the speed of
sound, with opposing influence on the instability growth. Moreover, as $\gamma$ or
the speed of sound (varied by changing the equilibrium pressure $p_\infty$ at the
interface while keeping the interface equilibrium density constant) increase to $\infty$, 
the limiting incompressible flows (and the bases of comparison) are different. 
The compressible growth rate is bounded by the growth rates obtained for these 
two incompressible flows which have exponentially varying and constant density, 
respectively. 

A physical explanation for the decrease in the growth rate as $p_\infty$ decreases 
can be formulated based on the influence of $p_\infty$ on the local Atwood number.
As the interface develops, the heavier fluid reaches regions of larger and larger 
densities of the lighter fluid while the lighter fluid reaches regions of smaller 
densities of the heavier fluid. The local Atwood number in these regions away from 
the initial position of the interface depends on $p_\infty$, since the equilibrium 
density profile depends on $p_\infty$. Using the equilibrium density provided below, 
it can be shown that the local Atwood number is lower for the points on the interface 
above the initial position, while it is higher below, compared to a system at higher 
$p_\infty$ (less compressible). However, the decrease in $p_\infty$ leads to a larger
change in the local Atwood number above the initial position of the interface. 
The overall effect would be a decrease in the average Atwood number, thus offering an 
intuitive argument for the decrease of the growth rate with decreasing $p_\infty$. 
Nevertheless this argument might break for small domain sizes and the validity of the 
results for finite size domains is discussed. The influence of surface tension and 
viscosity on the growth rate are also considered.

\section{Linearized Equations}

The case of two superposed compressible ideal fluids separated by an 
interface at $x_1=0$ is considered. The fluids are subject to a constant 
gravitational acceleration $\mbox{\boldmath{$g$}}=(-g,0,0)$. For each fluid,
the motion is governed by the continuity, momentum transport, and energy 
transport equations

\begin{equation}
\frac{\partial \rho}{\partial t}+\frac{\partial\rho u_k}{\partial x_k}=0
\label{eqn1} 
\end{equation}

\begin{equation}
\frac{\partial \rho u_i}{\partial t}+\frac{\partial \rho u_i u_k}
{\partial x_k}=-\frac{\partial p}{\partial x_i}+\frac{\partial \tau_{ik}}
{\partial x_k}-\rho g \delta_{i1} 
\label{eqn2} 
\end{equation}

\begin{equation}
\frac{\partial \rho e}{\partial t}+\frac{\partial \rho e u_k}
{\partial x_k}=-p\frac{\partial u_k}{\partial x_k}+\tau_{jk}
\frac{\partial u_j}{\partial x_k}+\frac{\partial }{\partial x_k}\left(
\lambda\frac{\partial T}{\partial x_k}\right),
\label{eqn3a} 
\end{equation}

\noindent
where $\rho$ is the density, $u_i$ the velocity in $x_i$ direction, $p$ the
pressure, $e$ the specific internal energy, and $T$ the temperature. The viscous
stress is assumed Newtonian, $\tau_{ij}=\mu\left(\frac{\partial u_i}
{\partial x_j}+\frac{\partial u_j}{\partial x_i}-\frac{2}{3}\frac{\partial 
u_k}{\partial x_k}\delta_{ij}\right)$, and the kinematic viscosity, $\mu$, 
and thermal conduction coefficient, $\lambda$, are considered constant. The
equations \ref{eqn1}-\ref{eqn3a} should be supplemented with equations of 
state for the pressure and internal energy. For each fluid, the specific heats are
assumed constant, $p=R\rho T$ and $e=c_v T$, where $R$ is the gas constant and
$c_v=R/(\gamma-1)$ is the specific energy at constant volume. With these 
assumptions $e=\frac{p}{\rho(\gamma-1)}$ and the energy equation becomes

\begin{equation}
\frac{\partial p}{\partial t}=-\gamma p\frac{\partial u_k}{\partial x_k}
-u_k\frac{\partial p}{\partial x_k}+(\gamma-1)\tau_{jk}\frac{\partial u_j}
{\partial x_k}+(\gamma-1)\frac{\partial }{\partial x_k}\left(
\lambda\frac{\partial T}{\partial x_k}\right).
\label{eqn3}
\end{equation}

\subsection{``0''-order equations}

The two fluids are assumed initially at rest and the primary variables
are written as small perturbations around the equilibrium (``0'') state,
with $\mbox{\boldmath{$u$}}\equiv 0$. Then the governing equations reduce to

\begin{eqnarray}
\frac{\partial \rho_0}{\partial t}&=&0\\
\frac{\partial p_0}{\partial x_1}&=&-\rho_0g,\mbox{  }\frac{\partial p_0}
 {\partial x_2}=\frac{\partial p_0}{\partial x_3}=0\\
\frac{\partial p_0}{\partial t}&=&(\gamma-1)\frac{\partial }{\partial x_1}
 \left(\lambda\frac{\partial T_0}{\partial x_1}\right).
\end{eqnarray}

\noindent
Moreover, the ``0''$^{th}$-order variables are assumed to be in steady state,
so that $\frac{\partial p_0}{\partial t}=0$ and $T_0$ and $p_0$ are
continuous across the interface. For infinite domain or finite size domain
in $x_1$ direction with adiabatic walls, the energy equation yields
$T_0=constant$. Consequently, the equilibrium state is given by

\begin{eqnarray}
\rho_{0_m}&=&\frac{p_\infty}{R_m T_0}\exp(-\frac{g}{R_mT_0}x_1)\\
p_{0_m}&=&p_\infty\exp(-\frac{g}{R_mT_0}x_1)\\
T_0&=&constant,
\end{eqnarray}

\noindent
where $p_\infty$ is the unperturbed pressure at the interface ($x_1=0$)
and $m=1,2$ denotes the material $1$ or $2$, with material $2$ chosen to be
above material $1$.

\subsection{First-order equations}

The interface between the two fluids is perturbed with an $x_2$ and $x_3$ dependent
perturbation. The location of the interface is described by the function 
$x_s(x_1,x_2,x_3,t)$, with $\frac{\partial x_s}{\partial t}=u_1$. Moreover, a 
surface tension is added at the interface between the two fluids. The first order 
linearized equations are obtained as

\begin{equation}
\frac{\partial \rho}{\partial t}+\rho_0\Delta+u_1D\rho_0=0
\label{lin1}
\end{equation}

\begin{eqnarray}
\rho_0\frac{\partial u_1}{\partial t}&=&-Dp-\rho g+
 \mu\frac{\partial}{\partial x_j}\left(\frac{\partial u_1}{\partial x_j}+
 Du_j\right)-\frac{2}{3}\mu D\Delta\nonumber\protect\\
&+&\sum_ST_s\left(\frac{\partial^2x_s}{\partial x_2\partial x_2}+
 \frac{\partial^2x_s}{\partial x_3\partial x_3}\right)\delta(x_1-x_s)
\label{lin2}
\end{eqnarray}

\begin{equation}
\rho_0\frac{\partial u_2}{\partial t}=-\frac{\partial p}{\partial x_2}+
\mu\frac{\partial}{\partial x_j}\left(\frac{\partial u_2}{\partial x_j}+
\frac{\partial u_j}{\partial x_2}\right)-\frac{2}{3}\mu\frac{\partial 
\Delta}{\partial x_2}
\label{lin3}
\end{equation}

\begin{equation}
\rho_0\frac{\partial u_3}{\partial t}=-\frac{\partial p}{\partial x_3}+
\mu\frac{\partial}{\partial x_j}\left(\frac{\partial u_3}{\partial x_j}+
\frac{\partial u_j}{\partial x_3}\right)-\frac{2}{3}\mu\frac{\partial 
\Delta}{\partial x_3}
\label{lin3a}
\end{equation}

\begin{equation}
\frac{\partial p}{\partial t}=-\gamma p_0 \Delta-u_1Dp_0+(\gamma-1)
 \frac{\partial}{\partial x_k}\left(\lambda\frac{\partial T}{\partial x_k}
\right),
\label{lin4}
\end{equation}

\noindent
where $\Delta=\frac{\partial u_j}{\partial x_j}$ is the dilatation and
$D$ denotes $\frac{\partial}{\partial x_1}$. Equations \ref{lin1}-\ref{lin4}
should be supplemented by boundary conditions and jump conditions across
the interface. These conditions will be discussed at length in the next
sections.

By examining equations \ref{lin1}-\ref{lin4} it should be noted that in the
absence of heat diffusion, the limit of incompressible flow ($\Delta=0$) can be obtained
either by letting $\gamma \rightarrow \infty$ (as considered by the previous 
authors) or by letting $p_\infty \rightarrow \infty$ (note that since the equilibrium
density at the interface should not be affected by the change in pressure, the equation of 
state implies $T_0\rightarrow \infty$ in this case). In the latter case the equilibrium 
density becomes $\rho_0=constant$, while in the former the exponential variation is still 
allowed. Therefore the incompressible limit (and the base of comparison for the rate of 
growth) is different in the two cases. In the next sections $\gamma$ and $p_\infty$ are 
considered as independent parameters, both affecting the compressibility of the flow. 
The inviscid, infinite domain and no surface tension case will be examined first. 
Then the influence of finite size domain, surface tension and viscosity on the results 
obtained will be investigated. The non-zero heat diffusivity case was also 
considered, but the results were very close to the non-diffusive case and are not 
presented here.

\section{Inviscid case}

In general, the linearized equations do not admit analytical solutions.
However, it is possible to obtain an analytical solution in the absence of
viscosity and heat diffusion. Following the usual approach (e.g. Ref. \cite{CH81}),
we seek solutions whose dependence on $x_2$, $x_3$, and time have the form

\begin{equation}
\exp(\mbox{\boldmath{$i$}}(k_2x_2+k_3x_3)+nt)
\end{equation}

\noindent
where $k_2$, $k_3$ and $n$ are constants. For solutions having this dependence,
equations \ref{lin1}-\ref{lin4} with $\mu=0$, $\lambda=0$, become

\begin{eqnarray}
n\rho&=&-\rho_0\Delta-u_1D\rho_0\\
\rho_0nu_1&=&-Dp-g\rho-k^2T_s\delta(x_1-x_s)\frac{u_1}{n}\\
\rho_0nu_2&=&-\mbox{\boldmath{$i$}}k_2p\\
\rho_0nu_3&=&-\mbox{\boldmath{$i$}}k_3p\\
np&=&-\gamma p_0\Delta +u_1\rho_0g,
\end{eqnarray}

\noindent
where $\Delta=Du_1+\mbox{\boldmath{$i$}}(k_2u_2+k_3u_3)$ and $k^2=k_2^2+k_3^2$. 
After eliminating $p$, $\Delta$, $u_2$, and $u_3$ from the above equations, an 
equation for $u_1$ is obtained as

\begin{eqnarray}
u_1D\left[\rho_0\frac{g/c^2}{k^2+n^2/c^2}\right]&-&D\left[\rho_0\frac{Du_1}
{k^2+n^2/c^2}\right]+\rho_0u_1+\frac{k^2}{n^2}T_s\delta(x_1-x_s)u_1\nonumber
\protect \\&-&
\frac{k^2g^2}{n^2c^2}\frac{u_1}{k^2+n^2/c^2}\rho_0-u_1\frac{gD\rho_0}{n^2}=0,
\label{equ1i}
\end{eqnarray}

\noindent
where $c=\sqrt{\gamma \frac{p_0}{\rho_0}}$ is the speed of sound. The jump
condition at the interface can be obtained by integrating equation 
\ref{equ1i} over an infinitesimal element of $x_1$ which includes the 
interface

\begin{equation}
u_s\delta\left[\rho_0\frac{g/c^2}{k^2+n^2/c^2}\right]-\delta\left[\rho_0
\frac{Du_1}{k^2+n^2/c^2}\right]+\frac{k^2}{n^2}T_su_s-u_s\frac{g}{n^2}
\delta\rho_0=0,
\label{equ1j}
\end{equation}

\noindent
where $\delta f=f_+-f_-$, with $f_+=f(x_s+0)$, $f_-=f(x_s-0)$, is the jump
of a quantity f across the interface. The subscript $s$ denotes the
value which a quantity, continuous at the interface, takes at $x_1=x_s$.

On each side of the interface, $c^2$ and $\frac{D\rho_0}{\rho_0}=-\frac{g}
{RT_0}$ are constant, so that equation \ref{equ1i} becomes

\begin{equation}
D^2u_{1_m}-\frac{\gamma_m g}{c_m^2}Du_1-\left(k^2+\frac{n^2}{c_m^2}+
\frac{(\gamma_m-1) g^2 k^2}{n^2c_m^2}\right)u_1=0,
\end{equation}

\noindent
with the solution of the form $u_{1_m}=A_m\exp(\lambda_{1_m}x_1)+B_m
\exp(\lambda_{2_m}x_1)$, where 

\begin{equation}
\lambda_{1,2_m}=\frac{\gamma_m g}{2c_m^2}\pm k\sqrt{1+\frac{n^2}{k^2c_m^2}+
\frac{(\gamma_m-1)g^2}{n^2c_m^2}+\frac{\gamma_m^2g^2}{4k^2c_m^4}}.
\label{lameq}
\end{equation}

\noindent
Formula \ref{lameq} for $\lambda_{1,2_m}$ was also obtained by Amala. \cite{Amala95}
The coefficients $A_m$ and $B_m$ can be determined to a multiplying constant
from the conditions that $u_1$ vanishes at the rigid boundaries located at 
$x_1=-l_1$ and $x_1=l_2$ and that it is continuous over the interface. 
After replacing $u_1$ in the jump condition \ref{equ1j}, a dispersion 
relation can be obtained as

\begin{eqnarray}
\overline{n}^2&=&F_1F_2\left[\alpha_2\gamma_2(\gamma_1+\overline{n}^2M
 \alpha_1)-\alpha_1\gamma_1(\gamma_2+\overline{n}^2M\alpha_2)-\overline{T_s}
(\gamma_1+ \overline{n}^2M\alpha_1)(\gamma_2+\overline{n}^2M\alpha_2)\right]
\nonumber\protect\\
&/&\left[\alpha_1\gamma_1(\gamma_2+\overline{n}^2M\alpha_2)
 [\overline{\lambda}_{1_1}\exp(\overline{\lambda}_{1_1}L_1)-
 \overline{\lambda}_{2_1}\exp(\overline{\lambda}_{2_1}L_1)]F_2 
\right.\nonumber\protect\\
&-&\left. \alpha_2\gamma_2(\gamma_1+\overline{n}^2M\alpha_1)
 [\overline{\lambda}_{1_2}\exp(-\overline{\lambda}_{1_2}L_2)-
 \overline{\lambda}_{2_2}\exp(-\overline{\lambda}_{2_2}L_2)]F_1\right],
\label{disp}
\end{eqnarray}

\noindent
where the nondimensional quantities are defined as $\overline{n}^2=
\frac{n^2}{kg}$, $M=\frac{g(\rho_1+\rho_2)}{k p_\infty}$, $\overline{T}_s=
\frac{T_sk^2}{g(\rho_1+\rho_2)}$, $L_m=kl_m$, $\overline{\lambda}_{1,2_m}=
\frac{\lambda_{1,2_m}}{k}=\frac{\alpha_m M}{2}\pm \sqrt{1+\frac{1}{\gamma_m}
\overline{n}^2\alpha_mM+\frac{\gamma_m-1}{\gamma_m}\frac{\alpha_m M}
{\overline{n}^2}+\frac{\alpha_m^2M^2}{4}}$, $F_1=
\exp(\overline{\lambda}_{1_1}L_1)-\exp(\overline{\lambda}_{2_1}L_1)$, and 
$F_2=\exp(-\overline{\lambda}_{1_2}L_2)-\exp(-\overline{\lambda}_{2_2}L_2)$.
The densities $\rho_1=\frac{p_\infty}{R_1T_0}$ and $\rho_2=\frac{p_\infty}
{R_2T_0}$ are the values of $\rho_0$ on the two sides of the interface and
$\alpha_m=\rho_m/(\rho_1+\rho_2)$.

Relation \ref{disp} represents a generalization of the dispersion
relations obtained by previous authors. It includes the effect of a finite 
size domain and surface tension, and makes no isentropic assumption.
For $L_1,L_2\rightarrow \infty$ the dispersion relation becomes

\begin{equation}
\overline{n}^2=\frac{\alpha_2\gamma_2(\gamma_1+\overline{n}^2M\alpha_1)-
\alpha_1\gamma_1(\gamma_2+\overline{n}^2M\alpha_2)-\overline{T_s}
(\gamma_1+\overline{n}^2M\alpha_1)(\gamma_2+\overline{n}^2M\alpha_2)}
{\alpha_1\gamma_1(\gamma_2+\overline{n}^2M\alpha_2)
\overline{\lambda}_{1_1}-\alpha_2\gamma_2(\gamma_1+\overline{n}^2M\alpha_1)
\overline{\lambda}_{2_2}}.
\end{equation}

As explained above, the incompressible flow limit ($\Delta=0$) can be 
obtained either by letting $p_\infty\rightarrow \infty$ or 
$\gamma \rightarrow \infty$. As $p_\infty\rightarrow \infty$, the flow 
approaches incompressible flow with constant density, for which the 
nondimensional rate of growth is given by 

\begin{equation}
\overline{n'}_i^2=\frac{\alpha_2-\alpha_1-\overline{T_s}}{\alpha_1\coth(L_1)+
\alpha_2\coth(L_2)},
\label{nip}
\end{equation}

\noindent
which for infinite domain becomes $\overline{n'}_i^2=\alpha_2-\alpha_1-
\overline{T_s}$. \cite{CH81} On the other hand, as $\gamma \rightarrow \infty$,
the equilibrium flow has still exponential varying density, with the 
dispersion relation

\begin{eqnarray}
\overline{n''}_i^2&=&F_1F_2\left[\alpha_2-\alpha_1-\overline{T_s}\right]/
 \left[\alpha_1[\overline{\lambda^i}_{1_1}\exp(\overline{\lambda^i}_{1_1}
 L_1)-\overline{\lambda^i}_{2_1}\exp(\overline{\lambda^i}_{2_1}L_1)]F^i_2
 \right.\nonumber\protect\\
 &-&\left.\alpha_2[\overline{\lambda^i}_{1_2}
 \exp(-\overline{\lambda^i}_{1_2}L_2)-\overline{\lambda^i}_{2_2}
 \exp(-\overline{\lambda^i}_{2_2}L_2)]F^i_1\right]F^i_1F^i_2,
\label{nipp}
\end{eqnarray}

\noindent
where $\overline{\lambda^i}_{1,2_m}=\frac{\alpha_m M}{2}\pm\sqrt{1+
\frac{\alpha_mM}{\overline{n''}_i^2}+\frac{\alpha_m^2M^2}{4}}$ and $F^i_m$
are defined using $\overline{\lambda^i}_{1,2_m}$. The parameter $M$ is related 
to the exponent in the formula for the unperturbed density $\rho_{0_m}=\rho_m
\exp(-M\alpha_mkx_1)$. For infinite domain and no surface tension equation 
\ref{nipp} reduces to the formula derived by Bernstein and Book \cite{BB83}
 
\begin{equation}
\overline{n''}_i^2=-M\alpha_1\alpha_2+\sqrt{M^2\alpha_1^2\alpha_2^2+
 (\alpha_2-\alpha_1)^2}
\label{ninc2}
\end{equation}

\noindent
and it is easy to show that $\overline{n''}_i^2<\overline{n'}_i^2$ for $M>0$
and $\overline{n''}_i^2\rightarrow \overline{n'}_i^2$ as $M\rightarrow 0$. 

In the dispersion formula for the compressible case (equation \ref{disp}) 
the decrease in $p_\infty$ is equivalent to an increase in $M$ at $\gamma$ 
constant. Thus, for the compressible case, $M$ represents a measure of the 
compressibility effects on the rate of growth. On the other hand, $M$ is proportional 
to the ratio between the wavelength of the initial perturbation and the 
density exponential change length scale. For small values of $M$, this 
lengthscale is much larger than the wavelength of the initial perturbation, 
and the rate of growth approaches the incompressible, constant density result.

For infinite domain, an approximate relation for $\overline{n}^2$,
valid to order $O(M)$ for small values of $M$ is

\begin{eqnarray}
\overline{n}^2&\approx&\overline{n'}_i^2\left[1+M\left[\left(\frac{\alpha_1}
 {\gamma_1}-\frac{\alpha_2}{\gamma_2}\right)\alpha_1\alpha_2-
 \frac{\overline{T}_s}{2}\left(\frac{\alpha_1^2}{\gamma_1}+\frac{\alpha_2^2}
 {\gamma_2}\right)\right]+M\alpha_1\alpha_2\left(\frac{1}{\gamma_1}-\frac{1}
 {\gamma_2}\right)\right.\nonumber\protect\\
&-&\left. \frac{M}{2}\left(\frac{\gamma_1-1}{\gamma_1}\alpha_1^2-
 \frac{\gamma_2-1}{\gamma_2}\alpha_2^2\right)\right]-\frac{M}{2}\left(
 \frac{\gamma_1-1}{\gamma_1}\alpha_1^2+\frac{\gamma_2-1}{\gamma_2}
 \alpha_2^2\right).
\label{grate1}
\end{eqnarray}

\noindent
For $\gamma_1=\gamma_2=1$ the relation derived by Plesset and Prosperetti 
\cite{PP82} is recovered.
In general, for $\gamma_1,\gamma_2\geq 1$, it can be shown that relation 
\ref{grate1} implies $\overline{n}^2<\overline{n'}_i^2$ for any combination
of parameters, with the constraint $\rho_2>\rho_1$. Moreover, Fig. 1 
shows that $\overline{n}^2$ decreases as $p_\infty$ decreases. 
It should be noted that the same effects of decreasing $p_\infty$ on the 
nondimensional growth rate can be obtained either by increasing $g$ or 
decreasing $k$, with all other parameters kept constant. In other words, 
compressibility effects are more important at larger values of $g$ and 
lower wavenumbers, which is supported by the numerical solutions of the 
dispersion relation presented in Fig. 1. 

For large $M$ and no surface tension, it follows from the dispersion relation that

\begin{equation}
\overline{n}^2\approx\frac{1}{M}\frac{(\alpha_2-\alpha_1)^2}
{\alpha_1\alpha_2\left(\alpha_1\frac{\gamma_2-1}{\gamma_2}+\alpha_2
\frac{\gamma_1-1}{\gamma_1}+\sqrt{\alpha_1^2+\alpha_2^2+2\frac{\gamma_1
\gamma_2-2\gamma_1-2\gamma_2+2}{\gamma_1\gamma_2}\alpha_1\alpha_2}\right)}.
\end{equation}

\noindent
A further reduction in the rate of growth can be obtained by increasing
the adiabatic exponents. At the limit, when $\gamma_1,\gamma_2\rightarrow
\infty$, the nondimensional rate of growth becomes

\begin{equation}
\overline{n}^2\approx\frac{1}{M}\frac{(\alpha_2-\alpha_1)^2}{2},
\end{equation} 

\noindent
which is the same as that obtained from formula \ref{ninc2} for large $M$ by 
Bernstein and Book \cite{BB83} and Turner. \cite{Turner02} In general, for finite 
values of $M$ it can be shown \cite{BB83,Turner02} that increasing the ratio of 
specific heats leads to a decrease in the rate of growth, also supported by Fig.
1. However, the rate of growth has different sensitivities to the change 
of $\gamma_1$ and $\gamma_2$. Thus, as Fig. 1 shows, the change in the 
ratio of specific heats of the lower fluid leads to a larger change of $n$.
Therefore, the rate of growth is more sensitive to the change in compressibility of 
the lower fluid. However, as Fig. 1 indicates, these results are also 
sensitive to the value of the Atwood number. For large values of the Atwood number
the results obtained for the compressible cases show little sensitivity to changes
in the ratios of specific heats and the rate of growth obtained for the compressible 
case is close to the incompressible variable density result. Nevertheless, at small 
Atwood numbers and for large $M$, the values of $\gamma_1$ and $\gamma_2$ become
important in determining the rate of growth. In addition, the relative difference
between the growth rates obtained for the two limiting incompressible flows increases
as the Atwood number decreases, so that compressibility effects are larger at small 
Atwood numbers.

In conclusion, the instability growth rate for a compressible flow ($n$) in the 
inviscid, infinite domain and no surface tension case is bounded by the growth 
rates of the corresponding incompressible flows obtained for uniform ($n_i'$) and 
exponentially varying density ($n_i''$), so that $n_i''<n<n_i'$. As $p_\infty$ 
increases (so that the flow becomes less compressible), $n$ increases towards 
$n_i'$, while as $\gamma$ increases (so that the fluid becomes less compressible),
$n$ decreases towards $n_i''$. 
  
\subsection{Influence of Finite Size Domain}

For the case in which the domain is bounded by rigid boundaries located at
$x=-l_1$ and $x=l_2$, the growth rates obtained for constant density
incompressible flow and incompressible flow with exponentially varying density
are given by equations \ref{nip} and \ref{nipp}, respectively. Figure 2(a)
presents the growth rate as a function of the nondimensional parameter
$M$ for different domain sizes. For domain sizes not very small compared to the
wavelength of the perturbation, the nondimensional growth rate is still bounded
by $\overline{n}_i'$ and $\overline{n}_i''$ and the rate of growth decreases as
the domain size decreases. However, similar to the incompressible case, the 
decrease in $\overline{n}$ is less significant when $L_1$ is decreased. 
Therefore, for $L_1<L_2$ the growth rate varies more when $p_\infty$ is 
changed, so it is more sensitive to the change in compressibility.

For the extreme case when $L_2<<1$ (domain size small compared to the wavelength
of the initial perturbation) and $\gamma_1\approx 1$ it is possible,
as Fig. 2(b) shows, that the compressible growth rate becomes larger 
than the constant density incompressible growth rate for $M$ smaller than a 
critical value. Numerical solutions of the dispersion relation \ref{disp} for 
a large range of parameters indicate that the curve $\overline{n}^2$ can 
intersect only once the line $\overline{n}^2=\overline{n'}^2_i$. Therefore, an
analytical condition for the existence of the overshoot can be found by letting
$M\rightarrow 0$ (in which case the dispersion relation simplifies considerably)
and imposing $\overline{n}^2>\overline{n'}^2_i$. After some algebra one obtains

\begin{eqnarray}
&&\frac{\alpha_1\coth\ L1+\alpha_2\coth\ L_2}{2}\left[\frac{\gamma_1-1}
{\gamma_1}\frac{\alpha_1^2}{\overline{n'}^2_i}\Phi_1+At+2\alpha_1\alpha_2
\left(\frac{\alpha_1}{\gamma_1}-\frac{\alpha_2}{\gamma_2}\right)\right.
\nonumber\protect\\
&&\left.\frac{\coth\ L_1+\coth\ L_2}{\alpha_1\coth\ L1+\alpha_2\coth\ L_2}+
\frac{\gamma_2-1}{\gamma_2}\frac{\alpha_2^2}{\overline{n'}^2_i}\Phi_2\right]+\frac{At}{2}
\left(\frac{\alpha_1^2}{\gamma_1}\Phi_1+\frac{\alpha_2^2}{\gamma_2}\Phi_2
\right)>0
\label{cond1}
\end{eqnarray}

\noindent
where $\Phi_m=L_m\coth^2\ L_m-\coth\ L_m -L_m$ varies from $0$ to $-1$ as
$L_m$ increases from $0$ to $\infty$. Consistent with the numerical results,
condition \ref{cond1} can be fulfilled only for small values of the domain size
of the upper fluid and ratio of specific heats close to $1$ for the lower fluid. 

\subsection{Influence of Surface Tension}

The presence of surface tension tends to inhibit the growth of the instability.
Moreover, for the incompressible case there is a critical wavenumber
$k_c=[(\rho_2-\rho_1)g/T]^{1/2}$, \cite{CH81} so that the arrangement is stable
for $k>k_c$. By imposing $n=0$ in the dispersion relation \ref{disp} it can be 
seen that the critical wavenumber remains the same as in the incompressible 
case. However, for $T_s\neq0$, the
wavenumber appears as an explicit parameter in the dispersion relation 
\ref{disp} so the variations of $p_\infty$ and $k$ are no longer equivalent. 
Nevertheless, at each $k$, the nondimensional compressible rate of growth 
obtained for infinite domain is still bounded by $\overline{n'}^2_i$ and
$\overline{n''}^2_i$, as Fig. 3 shows. However, the lower limit is 
approached differently as $\gamma_1$ or $\gamma_2$ increase to $\infty$. Thus, the 
variation in the compressibility of the lower fluid is more important at lower
wavenumbers, while the variation in the compressibility of the upper fluid is
more important at higher wavenumbers. 

\section{Effect of Viscosity}

Consider the case of two viscous fluids, bounded by two rigid surfaces at 
$x=-l_1$ and $x=l_2$. Following the previous procedure, the linearized equations
can be reduced to a single fourth order ordinary differential equation in $u_1$
of the form

\begin{equation}
A_4 D^4 u_1+A_3 D^3 u_1+A_2D^2u_1+A_1Du_1+A_0u_1=0,
\label{visceq}
\end{equation}

\noindent
where the coefficients $A_i$ are given in the Appendix for the compressible
case and the two incompressible limiting cases. However, only for the uniform 
density incompressible case this equation has constant coefficients and an 
easily derived analytical solution. The boundary conditions for equation 
\ref{visceq} are $u_i=0$ at $x=-l_1$ and $x=l_2$, $u_i$ and tangential viscous
stresses continuous at the interface, and a jump condition found from the 
integration of the governing equation over the interface. For the compressible 
case, the condition that tangential velocities vanish at the rigid boundary can be 
written as $\Delta-Du_1=0$ at $x=-l_1$ and $x=l_2$, while the continuity of $u_2$ 
and $u_3$ at the interface leads to the continuity of $\Delta-Du_1$. The divergence
of the velocity fluctuations is given by

\begin{equation}
\Delta=B_3D^3u_1+B_2D^2u_1+B_1Du_1+B_0u_1,
\label{delta}
\end{equation}

\noindent
with the coefficients $B_i$ given in the Appendix. The continuity of the 
tangential viscous stress over the interface can be written as

\begin{equation}
\delta(\mu[D\Delta -D^2u_1-k^2u_1])=0.
\end{equation} 

\noindent
An expression for $D\Delta$ in terms of the derivatives of $u_1$ is provided
in the Appendix. The jump condition at the interface can be found by eliminating
$p$, $u_2$, and $u_3$ from the momentum equation and integrating the resulting 
equation over an infinitesimal element of $x_1$ which includes the interface

\begin{eqnarray}
\delta[(-\rho+\frac{\mu}{n}D^2)(\Delta-Du_1)]&+&\frac{k^2}{n}\delta(\mu Du_1)=
 -\frac{k^2}{n^2}[g(\rho_2-\rho_1)-k^2T_s]u_s\nonumber\protect\\
&-&\frac{2k^2}{n}(\mu_2-\mu_1)(\Delta-Du_1)_s
\label{jump}
\end{eqnarray}

\noindent
For $\Delta=0$ condition \ref{jump} reduces to the condition derived for the
incompressible case in Ref. \cite{CH81} In the Appendix an expression for 
$D^2\Delta$ is provided. Since $u_1$ can be found only to a multiplying 
constant, the boundary conditions are supplemented with the specification of
$u_1$ or one of its derivatives at one point inside the domain. Then equation 
\ref{visceq} together with the boundary conditions described above form a closed 
set of equations from which $u_1$ on each side of the interface and the rate of 
growth $n$ can be determined. For all cases considered $l_1$ and $l_2$ are large 
compared to the wavelength of the initial perturbation so the configuration is 
close to the infinite domain case. Equation \ref{visceq} was integrated on each
side of the domain using a fourth order Runge-Kutta scheme. In order to determine 
$n$ and $u_1$ from the matching conditions at the interface, a multidimensional
secant method (Broyden's method) was employed. 

Figure 4 presents numerical solutions of the viscous linearized equations for
different Atwood numbers. Consistent with the previous results, the compressible 
rate of growth is bounded by the incompressible rates of growth obtained for 
uniform density and exponentially varying density cases. Moreover, its behavior
is similar to the well known incompressible constant density result. It has a peak
at some critical wavenumber and decreases towards zero as the wavenumber becomes 
large. However, the location of the peak is different compared to the 
incompressible case, with the highest difference at small Atwood numbers. Here 
we should note a qualitative difference between the results obtained for the 
constant density incompressible case and those obtained for the compressible and 
variable density incompressible cases. At small Atwood numbers, the critical 
wavenumber decreases with Atwood number for the constant density incompressible 
case, while it increases for the other two cases. Moreover, as shown in the previous 
sections, the largest differences in the inviscid rate of growth compared to the 
constant density incompressible case are obtained at small wavenumbers and small
Atwood numbers. It is expected then that, for the viscous case at small Atwood numbers, 
the relative difference between the rate of growth obtained for the compressible and 
incompressible cases considered will be largest, also confirmed by Figs. 4(a) and 4(b). 
Although this difference should become very small as $M$ approaches zero, Fig. 4(b) 
shows that, for small Atwood numbers, it persists at smaller values of $M$. 

\section{Conclusions}

The effects of compressibility on the growth rate of Rayleigh-Taylor instability 
between two immiscible ideal fluids are examined in the linear regime. The results 
distinguish between the stabilizing and destabilizing character of
compressibility. For infinite domains, the growth rate $n$ obtained for the 
compressible case is bounded by the growth rates obtained for the corresponding
incompressible flows with constant and exponentially varying density, and this
result is not affected by the presence of surface tension or viscosity. For ideal
gases with zero heat diffusivity, the limiting incompressible flow 
(defined by $\partial u_i/\partial x_i=0$) can be attained either by increasing the 
ratio of specific heats, $\gamma$, or the speed of sound (varied by changing the 
equilibrium pressure at the interface at constant equilibrium density at the interface). 
The equilibrium density distribution for the limiting incompressible flow is different 
in the two cases. Moreover, the two parameters have opposing influence on the rate of 
growth. As the speed of sound is increased, the rate of
growth increases towards the value obtained for the corresponding constant density 
incompressible flow, while as $\gamma$ increases, $n$ decreases towards the value 
obtained for the corresponding incompressible flow with exponentially varying 
density. The presence of heat diffusion was also considered, but the results were very close 
to those obtained for the nondiffusive case and were not presented here.

The equilibrium density for a compressible flow varies exponentially with $x_1$ and
depends on $p_\infty$. Therefore, the local Atwood number changes as the interface 
moves away from the original position. Compared to a flow with a higher value of 
$p_\infty$ (less compressible), the local Atwood number decreases for the points on 
the interface situated above the initial position, while it increases for the points on the 
interface situated below the initial position. However, the change in the local 
Atwood number is larger above the initial position of the interface, so that the overall
effect would be a decrease of the average local Atwood number. This offers an
intuitive argument for the decrease of the growth rate as $p_\infty$ decreases. 
Moreover, this argument suggests that the bubble velocity decreases, while the spike
velocity increases for more compressible flows. On the other hand, as $\gamma$ 
decreases the fluids are more compressible, however the equilibrium density and 
pressure do not change. Therefore, as the heavier fluid moves towards regions of 
higher pressures, its volume decreases and the volume change is larger for more 
compressible fluids, so that the spike velocity decreases. Similarly, for more 
compressible fluids the bubble velocity increases. If the two fluids have different 
values for $\gamma$, it is shown that the growth rate is more sensitive to the change
in the ratio of the specific heats of the lower fluid. However, at large Atwood
numbers the rate of growth is little influenced by the values of $\gamma1$ and 
$\gamma2$ and $p_\infty$ becomes the main compressibility parameter. In addition,
it is shown that compressibility effects are more important at small Atwood numbers.

For domains bounded by rigid surfaces, the compressible growth rate is still 
bounded by the two incompressible growth rates described above, except for the 
extreme case when the domain size of the upper fluid  is small compared to the 
wavelength of the initial perturbation and $\gamma \approx 1$ for the lower fluid.
In this case, the compressible growth rate can become larger than the growth rate 
obtained for the corresponding constant density incompressible flow for values of 
the compressibility parameter $M=\frac{g(\rho_1+\rho_2)}{kp_\infty}$ smaller than a
critical value. An analytical condition for the existence of this overshoot is 
provided. In general, the results show that the compressible growth rate varies more
when the rigid boundary of the lower fluid is closer to the interface than the 
rigid boundary of the upper fluid, so that it is more sensitive to the change in 
compressibility.

The presence of surface tension tends to inhibit the growth rate of the 
instability and for the incompressible case there is a critical wavenumber above 
which the configuration becomes stable. It is shown that the value of this critical
wavenumber is not affected by compressibility. For wavenumbers below this critical 
value the general result presented above remains valid. However, the presence of 
surface tension modifies the sensitivity of the growth rate to a differential 
change in the value of $\gamma$ for the two fluids. At smaller wavenumbers, the 
change in $\gamma$ for the lower fluid is more important for the variation of $n$,
while the opposite holds true at higher wavenumbers. 

Numerical solutions of the linearized equations show that for viscous compressible
fluids, the growth rate behaves in a manner analogous to the incompressible growth
rate. It has a most unstable wavenumber and decreases towards zero at larger 
wavenumbers. 
Moreover, both the growth rate and the most unstable mode are bounded by the values
obtained for the corresponding constant and variable density incompressible flows.
For the constant density incompressible flow it is known that the most unstable 
mode moves to small wavenumbers as the Atwood number is decreased. The inviscid 
results presented in this paper show that the effects of compressibility are more 
important at small wavenumbers and small Atwood numbers. Consistent with these 
results, it is found that for viscous fluids compressibility becomes more important
at small Atwood numbers. For small enough Atwood numbers, the difference between
the compressible and incompressible growth rates will remain sizable at larger
values of the equilibrium pressure.

An interesting question raised by the results presented in this paper is 
if they remain valid in the nonlinear regime for single and/or multimode initial 
perturbation. Our preliminary numerical results seem to indicate a similar influence
of $p_\infty$ and $\gamma$ on the growth rate (and on the spike and bubble 
velocities) to that found in the linear regime. Moreover, even for large values of the
equilibrium pressure so that the early time results are close to the incompressible
flow results, the late time spike and bubble velocities become different than
in the incompressible case. Another interesting question is about the range of the amplitudes
of the perturbation for which the growth rate agrees with the linear theory prediction. 
Again, our preliminary numerical results seem to indicate that the range of validity of the 
linear assumption remains approximately the same as in the incompressible case. These 
results will be published elsewhere. 

This study was concerned with the effects of compressibility on the instability growth between
immiscible fluids with uniform equilibrium temperature. It does not cover many of the
configurations of practical interest, for example the presence of an equilibrium temperature 
gradient, a more general equation of state or diffuse interfaces, which might be important in 
certain applications. \cite{DHH62,K94} However this study offers a systematic approach for 
examining the effects of compressibility which could represent a 
starting point for analyzing different or more complex configurations.

\begin{acknowledgments}

The author would like to thank  Dr. Timothy T. Clark and Dr. Francis H. Harlow for 
their advice on the preparation of this article and many helpful and insightful 
discussions. Computational resources were provided through the Institutional Computing
Project, Los Alamos National Laboratory.

This work was supported by the U.S. Department of Energy.

\end{acknowledgments}

\appendix

\section{Equations for the viscous case}
\label{viscous}

Following the usual procedure for the incompressible, constant density case, 
the variables are nondimensionalized using $1/n_0=\left(\frac{g^2}
{\nu_\infty}\right)^{-1/3}$ as time scale and $1/k_0=
\left(\frac{g}{\nu_\infty^2}\right)^{-1/3}$ as lengthscale. The compressibility 
parameter $M$ is defined by $M=\frac{g(\rho_1+\rho_2)}{k_0p_\infty}$. For 
simplicity the kinematic viscous coefficient is considered continuous over the 
interface, so that $\mu_1/\rho_1=\mu_2/\rho_2$, with $\mu_1$ and $\mu_2$ constant 
on each side of the interface. The value of the kinematic viscous coefficient at
the interface is denoted by $\nu_\infty$.

The scaled equations for $u_1$ and $\Delta$ on each side of the interface can 
be written as

\begin{eqnarray}
A_4 D^4 u_1&+&A_3 D^3 u_1+A_2D^2u_1+A_1Du_1+A_0u_1=0
\label{viscu1}\\
\beta_1\Delta&=&B_3D^3u_1+B_2D^2u_1+B_1Du_1+B_0
\label{vdelta}
\end{eqnarray}

\noindent
where the coefficients (with the index $m$ denoting the side of the interface
suppressed for simplicity) are given by

\begin{eqnarray}
A_4&=&B_3\beta_1\beta_2\\
A_3&=&(DB_3+B_2)\beta_1\beta_2-B_3\omega\\
A_2&=&(DB_2+B_1)\beta_1\beta_2+\exp(\alpha M x)\beta_1^2-B_2\omega\\
A_1&=&(DB_1+B_0)\beta_1\beta_2-\frac{\beta_1^2}{n}-B_1\omega\\
A_0&=&DB_0\beta_1\beta_2-(n+k^2\exp(\alpha M x))\beta_1^2-B_0\omega\\
B_3&=&\frac{\exp(\alpha M x)}{\beta_2}\left(\gamma+\frac{4}{3}\beta_3\right)\\
B_2&=&-\frac{\alpha M\exp(\alpha M x)}{\beta_2^3}\left[\gamma-\left(2\gamma-
 \frac{4}{3}\right)\beta_3\right]\\
B_1&=&-\frac{n}{\beta_2^2}\left[\frac{\alpha^2M^2\exp(\alpha M x)}{n}\left(
 \gamma-1-\frac{1}{3}\beta_3\right)\right.\nonumber\protect\\
 &+&\left.\beta_2\left(\gamma+\frac{4}{3}\beta_3
 \right)\left(1+\frac{k^2}{n}\exp(\alpha M x)\right)\right]\\
B_0&=&-\frac{\alpha n M}{\beta_2^2}\left[\beta_3\left(\gamma-1-\frac{1}{3}
 \beta_3\right)+(2\gamma-1)\alpha M k^2\exp(\alpha M x)+\frac{M k^2\beta_2}
 {\alpha n^2}\right]
\end{eqnarray}
\begin{eqnarray}
DB_3&=&\frac{\gamma\alpha^2M^2n\exp(2\alpha M x)}{\beta_2^2}\\
DB_2&=&\frac{\alpha^3m^3n\exp(2\alpha M x)}{3\beta_2^3}\left(7\gamma-3-
 \frac{6\gamma-4}{3}\beta_3\right)\\
DB_1&=&\frac{\alpha M n}{\beta_2^3}\left[\alpha^3m^3\exp(2\alpha Mx)\left(3\gamma
 -2-\frac{1}{3}\beta_3\right)+\beta_3^2\beta_2\left(\frac{1}{3}-\gamma
 \frac{k^2}{\alpha M n^2}\right)+\beta_2^3\right]\\
DB_0&=&\frac{\alpha^2 M^2 n}{\beta_2^3}\left[\frac{\alpha^2M^2\exp(2\alpha M 
 x)}{n}(\gamma-1)^2\left(\gamma-1-\frac{1}{3}\beta_3\right)\right.\nonumber\\
&&\left.-(2\gamma-1)\alpha M k^2\exp(2\alpha M x)\left(\gamma-\frac{1}{3}
 \beta_3\right)+\frac{k^2}{\alpha M n^2}\beta_2^3\right]\\
\beta_1&=&\frac{n}{\beta_2^2}\left[\frac{\alpha^2M^2\exp(\alpha M x)}{n}
 (\gamma-1)\left(\gamma-1-\frac{1}{3}\beta_3\right)\right.\nonumber\protect\\
&&\left.-\beta_2^2\left(1+\gamma
 \frac{Mk^2}{\alpha n^2}+\frac{4k^2}{3 n}\exp(\alpha M x)\right)\right]\\
\beta_2&=&\gamma+\frac{1}{3}\beta_3\\
\beta_3&=&\alpha M n\exp(\alpha M x)
\end{eqnarray}

The equation for $D\Delta$ can be written as

\begin{equation}
D\Delta=\frac{\alpha M n}{\beta_2}\left[\frac{\gamma-1}{n}\Delta-\exp(\alpha M 
 x) D^2u_1+\frac{1}{n}Du_1+(n+k^2\exp(\alpha M x))u_1\right]
\label{ddelta}
\end{equation}

\noindent
while the equation for $D^2\Delta$ is

\begin{eqnarray}
D^2\Delta&=&\frac{\alpha M n}{\beta_2}\left[\frac{\alpha M(\gamma-1)}{n
 \beta_2}\left(\gamma -1-\frac{1}{3}\beta_3\right)\Delta-\exp(\alpha M x)D^3u_1
 \right.\nonumber\protect\\
&+&\left(\frac{1}{n}-\frac{(2\gamma-1)M\exp(\alpha M x)}{\beta_2}\right)D^2u_1
 \nonumber\protect\\
&&+\left(\frac{\alpha M}{n\beta_2}\left(\gamma -1-\frac{1}{3}\beta_3\right)+n+
 k^2\exp(\alpha M x)\right)Du_1\nonumber\protect\\
&+&\left.\left(\frac{\alpha M n}{\beta_2}\left(\gamma -1-\frac{1}{3}\beta_3
 \right)+\frac{(2\gamma-1)\alpha M k^2\exp(\alpha M x)}{n\beta_2}\right)u_1
 \right]
\label{d2delta}
\end{eqnarray}

For $\gamma\rightarrow \infty$ or $p_\infty\rightarrow \infty$ 
($M\rightarrow 0$), equations \ref{vdelta}, \ref{ddelta}, and \ref{d2delta} 
yield $\Delta=0$, $D\Delta=0$, and $D^2\Delta=0$, so the incompressible case is
recovered. In the case $\gamma= \infty$ the equation for $u_1$ simplifies to
  
\begin{eqnarray}
D^4u_1&-&(n\exp(-\alpha M x)+2k^2)D^2u_1+n \alpha M \exp(-\alpha M x)Du_1
 \nonumber\protect\\
&+& \left(n\exp(-\alpha M x)+\frac{\alpha M}{n}\exp(-\alpha M x)+k^2\right)k^2u_1=0
\label{vinc1eq}
\end{eqnarray}

If, furthermore, $M\rightarrow 0$ in equation \ref{vinc1eq}, then the well
known equation for uniform density incompressible fluid derived in Ref.\cite{CH81}
is obtained. The same equation can be obtained by letting $p_\infty\rightarrow
\infty$ ($M\rightarrow 0$) directly in equation \ref{viscu1}.

\newpage
\section*{Figure captions}

{\bf Fig. 1.} Nondimensional rate of growth as function of 
$\frac{1}{M}=\frac{kp_\infty}{g(\rho_1+\rho_2)}$ for different values of 
$\gamma_1$ and $\gamma_2$.

{\bf Fig. 2.} Finite size effect on the nondimensional rate of growth. 
(a) No symbols curves represent the compressible case with 
$\gamma_1=\gamma_2=1.4$, open symbols the corresponding constant density 
incompressible case and closed symbols the corresponding variable density 
incompressible case. (b) $L_1=0.1$, $L_2=0.05$. All cases have At=0.5.

{\bf Fig. 3.} Compressibility influence on the nondimensional rate of growth in the
presence of surface tension, for $\frac{T_s}{g(\rho_1+\rho_2)}=0.78\ m^2$. All 
compressible cases have $\frac{g(\rho_1+\rho_2)}{p_\infty}=4\ m^{-1}$ except the 
open symbol case for which $\frac{g(\rho_1+\rho_2)}{p_\infty}=0.4\ m^{-1}$.

{\bf Fig. 4.} Growth rate dependency on the the wave number for viscous fluids.
(a) The compressible and incompressible variable density cases have $M=0.1$.
(b) At=0.1, compressible and variable density incompressible cases showed with
thick lines correspond to $M=0.1$ and with thin lines to $M=0.01$.

\newpage
\mbox{\bf Fig. 1}
\vskip 3cm
\mbox{ }
\vskip 2cm
\begin{figure}[h]
\includegraphics[clip]{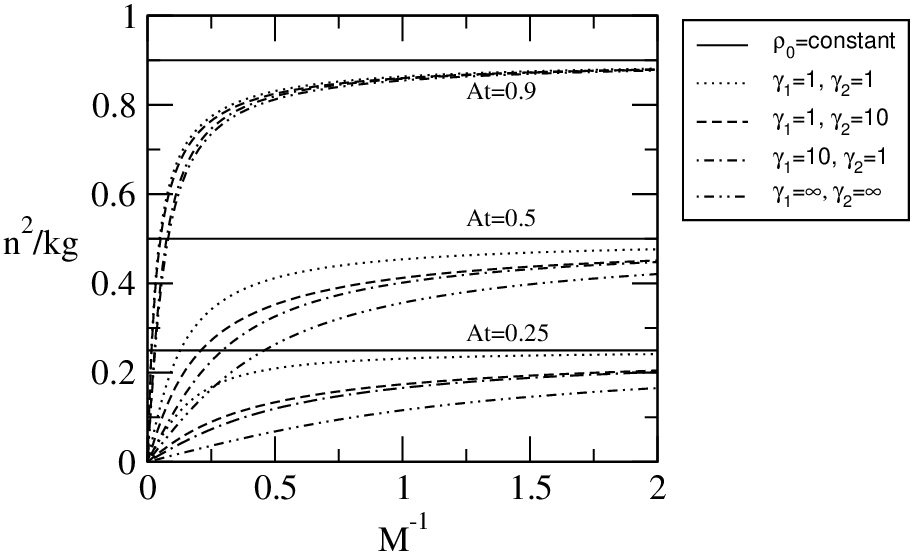}
\label{fig1}
\end{figure}

\newpage
\begin{minipage}[h]{2cm}
\mbox{\bf Fig. 2}
\vskip 1cm
\mbox{ }
\vskip 2cm
\end{minipage}
\begin{figure}[h]
\centerline{\includegraphics[height=8.5cm,clip]{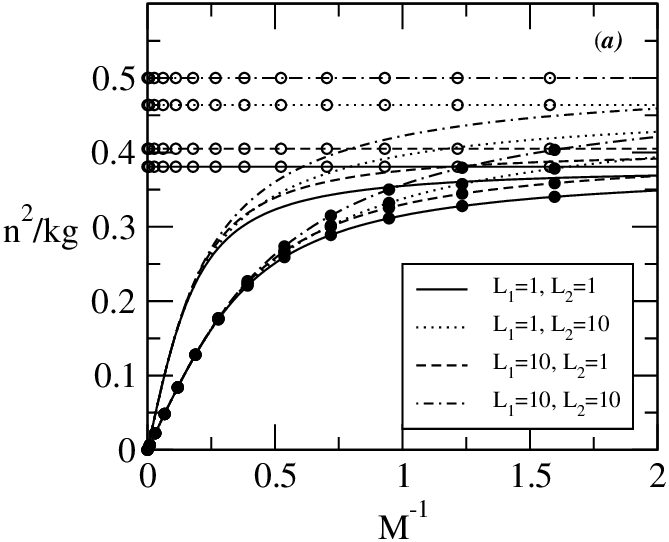}}
\end{figure}
\begin{figure}[h]
\centerline{\includegraphics[height=8.7cm,clip]{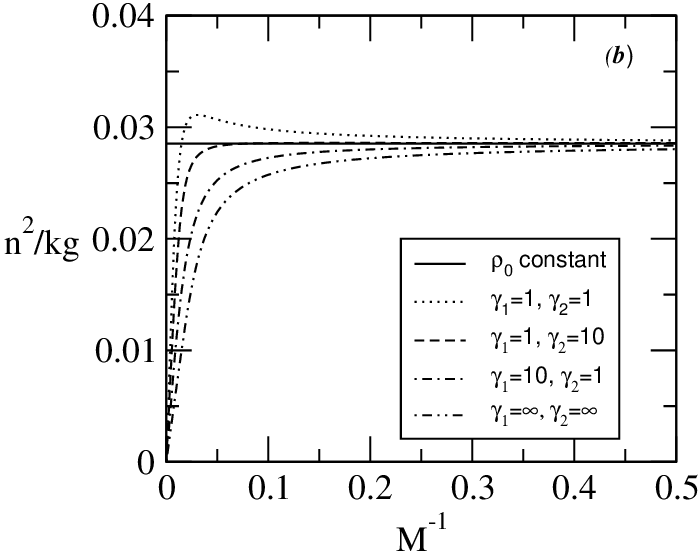}}
\label{fig2}
\end{figure}

\newpage
\mbox{\bf Fig. 3}
\vskip 3cm
\mbox{ }
\vskip 2cm
\begin{figure}[h]
\includegraphics{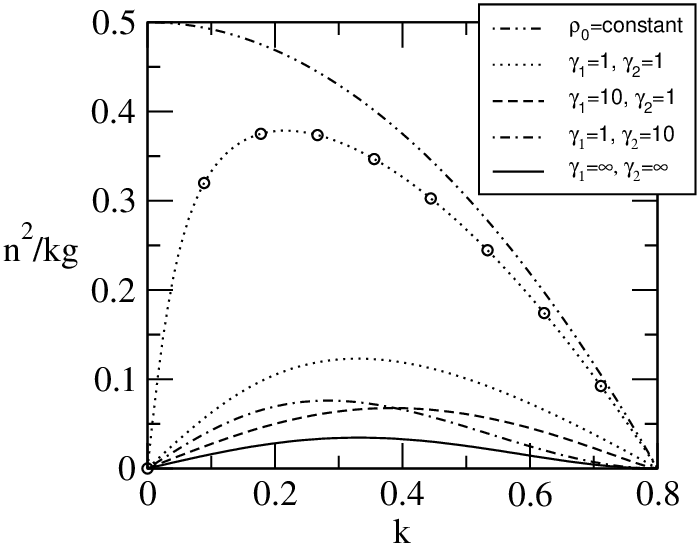}
\label{fig3}
\end{figure}

\newpage
\begin{minipage}[h]{2cm}
\mbox{\bf Fig. 4}
\vskip 1cm
\mbox{ }
\vskip 2cm
\end{minipage}
\begin{figure}[h]
\centerline{\includegraphics[height=8.5cm,clip]{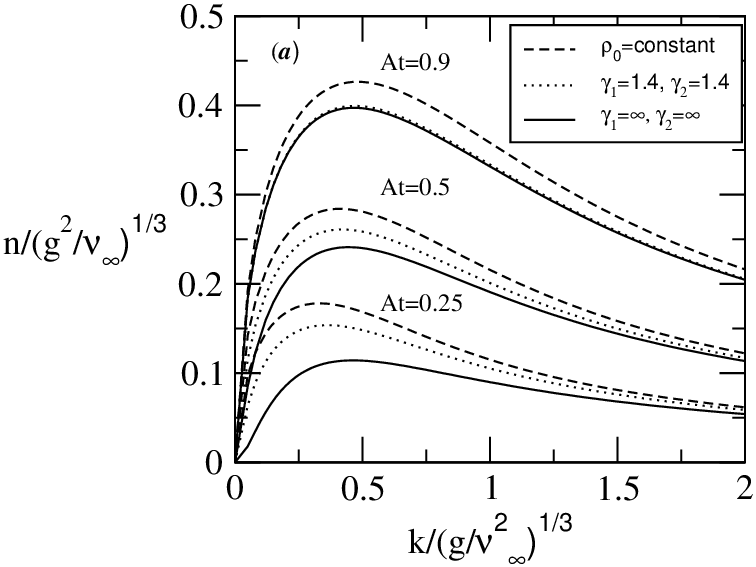}}
\end{figure}
\begin{figure}[h]
\centerline{\includegraphics[height=8.5cm,clip]{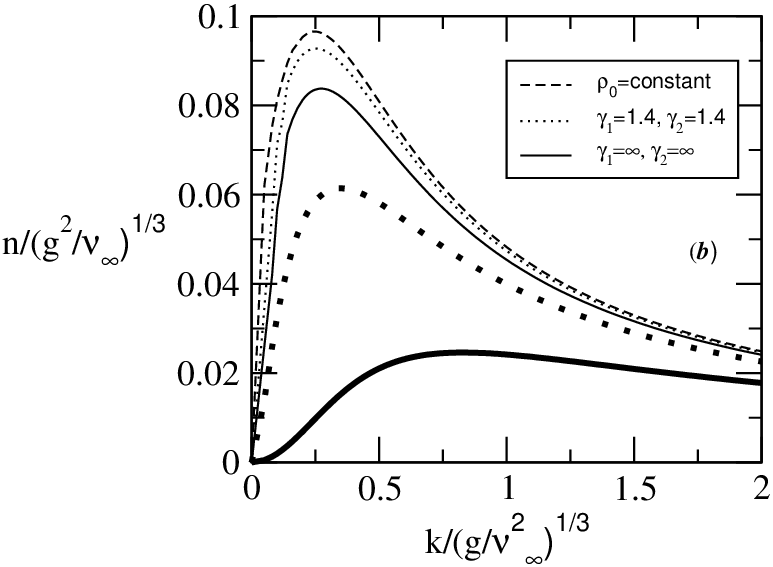}}
\label{fig4}
\end{figure}

\newpage

\end{document}